\documentstyle[12pt,epsf,epsfig,amsmath]{article}
\newfont{\thiplo}{msbm10 scaled\magstep 2}
\newfont{\gothic}{eufb10 scaled\magstep 2}
\newfont{\unc}{eurb10} 
\newskip\humongous \humongous=0pt plus 1000pt minus 1000pt
\def\caja{\mathsurround=0pt}
\def\eqalign#1{\,\vcenter{\openup1\jot \caja
        \ialign{\strut \hfil$\displaystyle{##}$&$
        \displaystyle{{}##}$\hfil\crcr#1\crcr}}\,}
\newif\ifdtup

\def\eqright #1\cr{\noalign{\hfill$\displaystyle{{}#1}$}}
\def\eqleft #1\cr{\noalign{\noindent$\displaystyle{{}#1}$\hfill}}

\def\oldreffmt#1{\rlap{[#1]} \hbox to 2\parindent{}}

\def\figfmt#1{\rlap{Figure {#1}} \hbox to 1in{}}

\def\sectioneq{\def\theequation{\thesection.\arabic{equation}}{\let
\holdsection=\section\def\section{\setcounter{equation}{0}\holdsection}}}%

\newcounter{holdequation}

\def\begineq #1\endeq{$$ \refstepcounter{equation}\eqalign{#1}\eqno
	(\theequation) $$}
\def\contlimit{\,{\hbox{$\longrightarrow$}\kern-1.8em\lower1ex
\hbox{${\scriptstyle (a\rightarrow0)}$}}\,}
\def\centeron#1#2{{\setbox0=\hbox{#1}\setbox1=\hbox{#2}\ifdim
\wd1>\wd0\kern.5\wd1\kern-.5\wd0\fi
\copy0\kern-.5\wd0\kern-.5\wd1\copy1\ifdim\wd0>\wd1
\kern.5\wd0\kern-.5\wd1\fi}}
\def\centerover#1#2{\centeron{#1}{\setbox0=\hbox{#1}\setbox
1=\hbox{#2}\raise\ht0\hbox{\raise\dp1\hbox{\copy1}}}}
\def\centerunder#1#2{\centeron{#1}{\setbox0=\hbox{#1}\setbox
1=\hbox{#2}\lower\dp0\hbox{\lower\ht1\hbox{\copy1}}}}
\def\lsim{\;\centeron{\raise.35ex\hbox{$<$}}{\lower.65ex\hbox
{$\sim$}}\;}
\def\gsim{\;\centeron{\raise.35ex\hbox{$>$}}{\lower.65ex\hbox
{$\sim$}}\;}

\def\super#1{\ifmmode \hbox{\textsuper{#1}}\else\textsuper{#1}\fi}
\def\textsuper#1{\newcount\holdspacefactor\holdspacefactor=\spacefactor
$^{#1}$\spacefactor=\holdspacefactor}

\def\getcite#1,{\advance\citenumber by1
\def\getcitearg{#1}\def\lastarg{@}
\ifnum\citenumber=1
\ref{#1}\let\next=\getcite\else\ifx\getcitearg\lastarg\let\next=\relax
\else ,\ref{#1}\let\next=\getcite\fi\fi\next}

\def\pom{{\rm P\kern -0.53em\llap I\,}}
\def\spom{{\rm P\kern -0.36em\llap \small I\,}}
\def\sspom{{\rm P\kern -0.33em\llap \footnotesize I\,}}

\relax
\def\contlimit{\,{\hbox{$\longrightarrow$}\kern-1.8em\lower1ex
\hbox{${\scriptstyle (a\rightarrow0)}$}}\,}
\def\upon #1/#2 {{\textstyle{#1\over #2}}}
\relax
\renewcommand{\thefootnote}{\fnsymbol{footnote}}

\sectioneq

\def\til#1{\centeron{\hbox{$#1$}}{\lower 2ex\hbox{$\char'176$}}}
\def\tild#1{\centeron{\hbox{$\,#1$}}{\lower 2.5ex\hbox{$\char'176$}}}
\def\sumtil{\centeron{\hbox{$\displaystyle\sum$}}{\lower
-1.5ex\hbox{$\widetilde{\phantom{xx}}$}}}

\begin{document}

\begin{titlepage} 

\rightline{\vbox{\halign{&#\hfil\cr
&FERMILAB-PUB-04-236-E\cr
&\today\cr}}} 
\vspace{0.25in}

\begin{center} 
  
{\large\bf  The White Pomeron, Color Sextet Quarks and Cosmic Ray Anomalies
}\footnote{(Talk given at 44th INFN Eloisatron Project Workshop;
QCD at Cosmic Energies, Erice, September 2004)}

\medskip

Michael G. Albrow

\vskip 0.6cm

\centerline{Fermi National Accelerator Laboratory,}
\centerline{Batavia, IL 60510, USA.}
\vspace{0.5cm}

\end{center}

\begin{abstract} 

 I introduce the critical pomeron theory (``white pomeron") proposed by Alan
  White, with its prediction of excessive $WW$ and $ZZ$ production at the LHC
  especially by double pomeron exchange (DPE), which is a crucial test. 
This theory implies dramatic effects in cosmic rays, possibly 
explaining the knee and high $E_T$ jet excesses. Some of the Ultra High 
Energy (UHE) cosmic rays may be
``Superbaryons" made of color sextet Quarks, and most (all?) UHE collisions in
  the atmosphere may produce $W$'s and $Z$'s.

\end{abstract} 

\renewcommand{\thefootnote}{\arabic{footnote}} 

\end{titlepage} 

At a Blois Workshop in 1989 Alan White~\cite{arw1} reviewed his ideas on
the critical pomeron and confinement, one of the implications being strong (QCD)
production of $W^+W^-$ and $ZZ$ pairs at sufficiently high energy.
At the same meeting I talked~\cite{mga1} on ``Double Pomeron Exchange from the ISR to the
SSC", referring to his prediction of ``... dramatic... direct and strong $W^+W^-$ 
and $ZZ$
production by pomerons..." at the SSC. The SSC is not to be, but the LHC should
have enough energy for this prediction to be well tested. A rule-of-thumb for
the masses $M_X$ which can be excited in DPE is 
$M_X \centerunder{$<$}{$\sim$} 0.05\sqrt{s}$ which
is 100 GeV at the Tevatron but 700 GeV at the LHC. Thus the LHC is comfortably
into the region where strong production of vector boson pairs by pomerons should be
manifest and could be dramatic. The Tevatron has probably too low $\sqrt{s}$,
but of course 100 GeV is just a rule of thumb and not a sharp cut-off, and given
that we begin to have data there it is certainly worth a look~\cite{arw2}.

In the first 200 pb$^{-1}$ of Run 2 data ($\sqrt{s} = 1960$ GeV) CDF has
approximately 20 $WW,WZ$ or $ZZ$ candidates. This number is consistent with
standard model expectations. Forward (high $x_F$) protons were not detectable
with high efficiency $\times$ acceptance,
so a search for a diffractive signature can only use the associated hadron
distributions for $|\eta| \centerunder{$<$}{$\sim$} 5.5$. This study is underway; some events look
strikingly clean but it is too early to say whether there is any significance to
this. A year from now we should have five times the data and a more
sophisticated Monte Carlo study (including the very forward region). If the
Tevatron data is inconclusive, the LHC should supply a definitive test.

The critical pomeron (critical meaning intercept = 1.0) plays a key role in
White's theory. It has two parts, a reggeized gluon and an infinite number of
``wee gluons". In QCD a leading order $qq$-scattering diagram by one-gluon
exchange is ``sick", not being gauge invariant. To meet this requirement one
adds other gluon exchanges and sums them up in a particular way to give a gauge
invariant sum, the reggeized gluon. This carries color and is not yet a pomeron.
In the BFKL pomeron the color is neutralized by a second reggeized gluon linked
to the first by the rungs of a ladder. In the ``white pomeron" the color is
neutralized at large distance by a cloud of wee gluons, gluons that have no (or very
small) momentum even in the infinite momentum frame. These wee gluons have the
properties of the vacuum. They are directly responsible for confinement and chiral
symmetry breaking in the theory and so, in a sense they {\em are} the vacuum.

Asymptotic freedom must be saturated to obtain the critical pomeron. This would require 16 color-triplet
quark flavors. Of course we only know 6 (d,u,s,c,b,t). However higher color
sextet Quarks would each count 5$\times$ in this sum, so a pair of such $Q_S$'s, say
\{$U,D$\}, will result in asymptotic freedom with the critical pomeron. The color
sextet quarks $Q_S$ have zero current mass, but a constituent mass at the
electroweak scale, probably a few hundred GeV. They have a stronger color charge
than triplet quarks $q$. They would have electric charge opposite to that of
quarks, i.e. $Q(U) = -\frac{2}{3}$. They can form ``Superhadrons" such as 
$P_6$ =[$UUD$] and $N_6$ = [$UDD$]. The $N_6$ is probably the lightest Superbaryon,
hence stable, with a mass at the TeV scale (within a factor ~2 ?), and would be a candidate
for dark matter. It is strongly interacting (a SIMP) through diffractive
processes, and could be a component of UHE cosmic rays detectable e.g. in AUGER.
At low energies, way below the $Q_S\bar{Q_S}$ threshold, it behaves like a WIMP.
Anomaly cancellation requires a pair of heavy Lepton doublets
 \{${L^1,\nu_6^1}$\} and \{${L^2,\nu_6^2}$\} or some other lepton sector. The $\nu_6$'s would be another (WIMP) 
component of dark matter.

Supermesons are formed of [$Q_S\bar{Q_S}$] giving e.g. \{$\Pi^-\Pi^\circ\Pi^+$\} and
$\eta_6$. The \{$\Pi^-\Pi^\circ\Pi^+$\} are composite zero-helicity components of
\{$W^-Z^\circ W^+$\} (they get ``eaten" by massless \{$W^-Z^\circ W^+$\} to give
them their masses). The $\eta_6$ plays the role of the Higgs boson in
electroweak symmetry breaking. There is no ``standard model" Higgs in this
theory! (The $\eta_6$ has rather different properties.) This form of symmetry
breaking is similar, in some respects, to higher representation technicolor and,
because of this, may be consistent with electroweak precision measurements. The
difference is that no new technicolor gauge fields are needed.

The wee gluon cloud in the pomeron couples particularly strongly to the color sextet quarks.
Their color charge is large and the number of gluons is large. And the color sextet
quarks couple strongly to $W$ and $Z$. Once above the
electroweak scale, and into the realm of $Q_S$'s, there should be prolific
diffractive
production of $W,Z$. At UHE cosmic ray energies (AGASA, AUGER)
they may be produced almost like pions, with several in most events! 

\begin{figure} [t]
\vspace{9.0cm}
\includegraphics{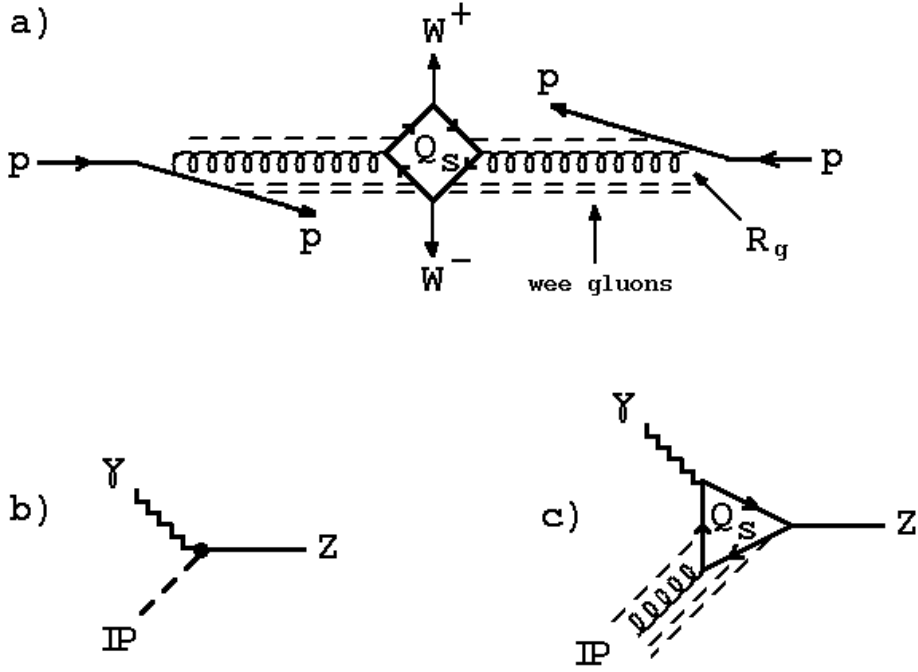}
\caption{\it
(a) Production of W-pairs by DPE through $Q_S$ loop 
(b) $\gamma I\!\!P Z$ vertex (c) $\gamma I\!\!P Z$
coupling through $Q_S$ loop. \label{diag} }
\end{figure}

A diagram of $W^+W^-$ pair production by DPE is shown in Fig.\ref{diag}. Fig 1b shows the striking 
$\gamma I\!\!P Z$ vertex through a sextet Quark loop (Fig 1c). This photoproduction of
$Z$ would make very high energy $ep$ collisions very interesting (ILC on a proton
storage ring: Tevatron, HERA or LHC). Pomeron physics is difficult at an $e^+e^-$
machine; how much of this sector can be addressed at the ILC itself ($e^+e^-$) has to be assessed.

Is there any evidence for the color sextet quark theory? To those who consider dark
matter evidence for SUSY it is also evidence for this theory which has dark matter
candidates! But of course we must first find dark matter particles. There are very
significant anomalies in the rate of high $E_T$ jets (``cores") in experiments such as
Chacaltaya and Kanbala~\cite{cao}. A QCD Monte Carlo was tuned to jet data at fixed
target and collider energies (including the $Sp\bar{p}S$ and Tevatron). For cosmic rays
with energies above $\sqrt{s} \approx $ 5 TeV the jet rate for 
$\chi_{12} \centerunder{$>$}{$\sim$} $ 1000
TeV.cm exceeds the data by an order of magnitude, where $\chi_{12}$ is basically the
product of the jets' $E_T$ and the jet-pair separation $R_{12}$. This seems to occur
between Tevatron and LHC energies. Could these jets actually be high $E_T$ $W$ and $Z$?
There are also indications of an excess of jets at high $E_T$ in CDF data, but the
situation is unclear as the systematic errors are large, and D0 does not seem to see the
same effect. It is also in a region where poorly known large-$x$ gluons could be
important. However there are (at least) two mechanisms giving such an excess in this
theory. They might be anomalous (strongly produced) $W,Z$'s, and this may be testable in existing data. They
might be due to a slowing down of the running of $\alpha_s$ as the scale $Q^2$
approaches the (non-perturbative) $Q_S$ scale. (This $\alpha_s$-slowing was
actually``seen" in the CDF Run-I data~\cite{cdfa} {\em under the assumption of the
CTEQ4M pdf}.) However I want to emphasize that this possible jet excess, while
intriguing, cannot at this stage be considered as evidence for {\em anything}. The
knee in the cosmic ray spectrum around $E_\circ \approx 3.10^6$ GeV, $\sqrt{s} \approx 2500$
GeV is however very well established and not understood. In White's theory this is about
the energy at which $W$'s and $Z$'s start to be strongly pair produced, so the transverse
profile of the atmospheric cascade broadens and more of the primary energy goes into
neutrinos, both effects leading to an underestimate of the primary energy and hence a
steepening of the spectrum~\cite{bora}.
 
Even though the Tevatron is probably at too low $\sqrt{s}$ to see evidence for the white
pomeron and its effects, I believe it is worth looking. The threshold would not be a
$\theta$-function. So far we have looked at only 200 pb$^{-1}$ for $WW/WZ/ZZ$ and do not
see a significant anomaly, though there are some intriguing events. Unfortunately we do
not have precision roman pots (on both outgoing beams and with good acceptance) to tag the $p$ and $\bar{p}$. If
we did one could even look for $Z \rightarrow \nu\bar{\nu}$ through missing mass,
measure $M(WW)$ etc. Technically it would be possible to install such pots, and we looked
into it, but it would require a few month shut-down, cost $\approx$ \$0.6M and disturb the
machine, so it would require a very strong case (5$\sigma$ signal?). However LHC should
be with us in 4 years, and CMS and perhaps ATLAS will have the ability to see the
diffractively scattered protons in $W^+W^-$ and $ZZ$ events. If ``White is right" there
will be dramatic effects.

\section{Acknowledgements}
I thank Alan White for many discussions and for comments on this talk.

\end{document}